\documentstyle[12pt]{article}
\begin{document}
\title{A BRIEF NOTE ON THE MAGNETIC EFFECTS OF THE ELECTRON}
\author{B.G. Sidharth$^*$\\ Centre for Applicable Mathematics \& Computer Sciences\\
B.M. Birla Science Centre, Hyderabad 500 063 (India)}
\date{}
\maketitle
\footnotetext{E-mail:birlasc@hd1.vsnl.net.in}
\begin{abstract}
In this paper it is shown that a recent formulation of the electron in terms
of a Kerr-Newman type metric, exhibits a short range magnetic effect, as
indeed has been observed at Cornell, and also an Aharonov-Bohm type of an
effect.
\end{abstract}
In a recent model\cite{r1,r2,r3} it was shown how an electron could be described
as a Kerr-Newman type black hole with Quantum Mechanical inputs. Such a scheme
lead to a cosmology consistent with all
so called large number relations and which predicted that the universe would
continue to expand for ever\cite{r4,r5}, as indeed has been subsequently
observed\cite{r6,r7}. Moreover this scheme also
gives a description of the quark picture including such features as the
characteristic fractional charge, handedness, confinement and an order of
magnitude estimate of the masses\cite{r8,r9,r10}.\\
We would now like to point out two additional consequences of the above model,
one an extra magnetic effect in the electromagnetic vacuum and the other an
Aharanov-Bohm type effect\cite{r11}.\\
We first observe that the magnetic component of the field of a static electron
as a Kerr-Newman black hole is given in the familiar spherical polar coordinates
by (Cf.refs.\cite{r1,r2})
\begin{equation}
B_{\hat r} = \frac{2ea}{r^3} cos \Theta + 0(\frac{1}{r^4}), B_{\hat \Theta} =
\frac{ea sin\Theta}{r^3} + 0(\frac{1}{r^4}), B_{\hat \phi} = 0,\label{e1}
\end{equation}
whereas the electrical part is given by
\begin{equation}
E_{\hat r} = \frac{e}{r^2} + 0(\frac{1}{r^3}), E_{\hat \Theta} = 0(\frac{1}{r^4}),
E_{\hat \phi} = 0,\label{e2}
\end{equation}
A comparison of (\ref{e1}) and (\ref{e2}) shows that there is a magnetic
component of shorter range apart from the dipole which is given by the first
term on the right in equation (\ref{e1})-- infact this model also exhibits the anomalous
gyro magnetic ratio $g=2$ of the electron. We would like to point out that a
short range force the $B^{(3)}$ force mediated by massive photons has
indeed been observed at Cornell and studied over the past few years\cite{r12}.\\
On the other hand as the Kerr-Newman charged black hole can be approximated
by a solinoid, we have as in the Aharonov-Bohm effect, a negligible magnetic
field outdside, but at the same time a real vector potential $\vec A$ which
would contribute to a shift in phase. Infact this shift in phase is given
by
\begin{equation}
\Delta \delta_{\hat B} = \frac{e}{\hbar} \oint \vec A . \vec{ds}\label{e3}
\end{equation}
There is also a similar effect due to the electric charge given by
\begin{equation}
\Delta \delta_{\hat E} = -\frac{e}{\hbar} \int A_0 dt\label{e4}
\end{equation}
where $A_0$ is the usual electro static potential given in (\ref{e2}). In the above Kerr-Newman
formulation, $(\vec A ,A_0)$ of (\ref{e3}) and (\ref{e4}) are given by (Cf.refs.\cite{r1,r2})
\begin{equation}
A_\sigma = \frac{1}{2}(\eta^{\mu v}h_{\mu v}),\sigma,\label{e5}
\end{equation}
From (\ref{e5}) it can be seen that
\begin{equation}
\vec A \sim \frac{1}{c} A_0\label{e6}
\end{equation}
Substitution of (\ref{e6}) in (\ref{e3}) then gives us the contribution of the
shift in phase due to the magnetic field.

\end{document}